\documentclass[aps,pra,twocolumn,showpacs,amssymb,amsmath]{revtex4-1}
\usepackage{graphics}
\usepackage{bm}
\usepackage{graphicx}
\usepackage{amsmath}
\usepackage{amssymb}
\usepackage{epsfig}
\usepackage{latexsym}
\usepackage{feynmf}
\begin{document}
\title{Entanglement of two harmonic modes coupled by angular momentum}\vskip2cm
\author{L.\ Reb\'on and R.\ Rossignoli}
\affiliation{{\small\it Departamento de F\'{\i}sica-IFLP, Universidad Nacional
de La Plata,}  {\small\it C.C.~67, 1900 La Plata, Argentina}}
\begin{abstract}
We examine the entanglement induced by an angular momentum coupling between two
harmonic systems. The Hamiltonian corresponds to that of a charged particle in
a uniform magnetic field in an anisotropic quadratic potential, or
equivalently, to that of a particle in a rotating quadratic potential. We
analyze both the vacuum and thermal entanglement, obtaining analytic
expressions for the entanglement entropy and negativity through the gaussian
state formalism. It is shown that vacuum entanglement diverges at the edges of
the dynamically stable sectors, increasing with the angular momentum and
saturating for strong fields, whereas at finite temperature, entanglement is
non-zero just within a finite field or frequency window and no longer diverges.
Moreover, the limit temperature for entanglement is finite in the whole stable
domain. The thermal behavior of the gaussian quantum discord and its difference
with the negativity is also discussed.
\end{abstract}
\maketitle

\section{Introduction}
The investigation of quantum entanglement and quantum correlations in distinct
physical systems is of great interest for both quantum information and
many-body physics \cite{NC.00,VE.07,AFOV.08,HHHH.10}. While the evaluation of
entanglement in systems with a high dimensional Hilbert space is in general a
difficult problem, boson systems described by quadratic Hamiltonians in the
basic boson operators offer the invaluable advantage of admitting an exact
evaluation of entanglement measures in both the ground and thermal state,
through the gaussian state formalism
\cite{RS.00,WW.01,AEPW.02,ASI.04,CEPD.06,AI.08,MRC.10}. The latter allows to
express the entanglement entropy \cite{BBPS.96} and negativity
\cite{ZHSL.99,VW.02} of any bipartition of a gaussian state in terms of the
symplectic eigenvalues of covariance matrices of the basic operators. Moreover,
the positive partial transpose (PPT) separability criterion \cite{PP.96,HHH.96}
is both necessary and sufficient for two-mode mixed gaussian states
\cite{RS.00} (and also $(1,n-1)$ bipartitions of $n$ modes gaussian states
\cite{WW.01}), turning the negativity into a rigorous entanglement indicator
for these systems. Let us also remark that there is presently a great interest
in continuous variable based quantum information \cite{BvL.05}, where gaussian
states constitute the basic element.

In addition, an approximate yet analytic evaluation of the {\it quantum
discord} \cite{OZ.01,HV.01} in two-mode gaussian states was recently achieved
\cite{GP.10,AD.10}, by restricting the local measurement that determines this
quantity to a gaussian measurement \cite{BvL.05}. Quantum discord is a measure
of quantum correlations which coincides with the entanglement entropy in pure
states but differs essentially from entanglement in mixed states, where it can
be non-zero even if the state is separable, i.e., with no entanglement. The
current interest in the quantum discord was triggered by its presence
\cite{DSC.08} in certain mixed state based quantum computation schemes which
provide exponential speedup over classical ones \cite{KL.98}, yet exhibiting no
entanglement \cite{DFC.05}. Important properties of states with non-zero
discord were recently unveiled \cite{CAB.11,SKD.11,PGA.11}.

The aim of this work is to examine, using the gaussian state formalism, the
entanglement and quantum correlations between two harmonic modes generated by
an angular momentum coupling. Such system arises, for instance, when
considering a charged particle in a uniform magnetic field in an anisotropic
quadratic potential, or also a particle in a rotating anisotropic harmonic trap
\cite{Va.56,FK.70,RS.80, BR.86}. The model has then been employed in several
areas, including the description of deformed rotating nuclei \cite{RS.80,BR.86}, 
anisotropic quantum dots in a magnetic field \cite{MC.94}, and fast rotating Bose-Einstein condensates \cite{LNF.01,OO.04,AF.07,ABL.09} in the
lowest Landau level approximation \cite{ABD.05,BDS.08,AF.09}. Containing just
quadratic couplings in the associated boson operators, the different terms in
the Hamiltonian may in principle be also simulated by standard optical means
\cite{BvL.05,PE.94}. For a general quadratic potential, the model exhibits a
complex dynamical phase diagram \cite{RK.09}, presenting distinct types of
stable and unstable domains and admitting the possibility of stabilizing an
initially unstable system by increasing the field or frequency \cite{RK.09}.
The model provides then an interesting and physically relevant scenario for
analyzing the behavior of mode entanglement in different regimes and near the
onset of different types of instabilities, with the advantage of allowing an
exact analytic evaluation of entanglement and quantum correlation measures at both zero and finite temperature. In addition, the present results indicate that mode entanglement can be easily controlled in this systems by modifying the field or frequency, suggesting a potential for quantum information applications. Let us finally mention that the dynamics of entanglement in other
two-mode systems were examined in \cite{HMM.03,NL.05,CN.08}.

In sec.\ \ref{II} we describe the model and derive the analytic expressions for
the vacuum entanglement entropy and the thermal negativity. The basic features
of the quantum discord are also discussed. The detailed behavior of
entanglement with the relevant control parameters is then analyzed in sec.\
\ref{III}, where we show that while vacuum entanglement diverges at the edges
of stable sectors, being correlated with the angular momentum, at finite
temperature entanglement is finite, and non-zero just within a finite field
window and below a {\it finite} limit temperature. A comparison between the
thermal behavior of the negativity and that of the gaussian quantum discord is
finally made, which indicates a quite different thermal response of these two
quantities, with the discord vanishing only asymptotically for
$T\rightarrow\infty$. Conclusions are finally drawn in \ref{IV}.

\section{Formalism\label{II}}\label{formalism}
\subsection{Model Hamiltonian}
We consider a system described by the Hamiltonian
\begin{eqnarray}
H=\frac{1}{2}(P_x^2+k'_x Q_x^2)+\frac{1}{2}(P_y^2+k'_y Q_y^2)
-\omega (Q_x P_y-Q_y P_x)\,,\label{H1}
\end{eqnarray}
which represents two harmonic modes coupled by an angular momentum term. Here
$Q_\mu$, $P_\mu$ stand for dimensionless coordinates and momenta 
($[Q_\mu,P_\nu]=i\delta_{\mu\nu}$, $[Q_\mu,Q_\nu]=[P_\mu,P_\nu]=0$). Eq.\
(\ref{H1}) arises, for instance, in the description of a particle of charge $e$
and mass $m$  in a general quadratic potential subject to a uniform magnetic
field, parallel to a principal axis of the potential. Denoting this axis as
$z$, such Hamiltonian reads
\begin{eqnarray}
{\cal H}
&=&\frac{(\bm{P}-e\bm{A}/c)^{2}}{2m}+
\frac{1}{2}(K_{x}{\cal Q}_x^{2}+K_{y}{\cal Q}_y^{2}+
K_{z}{\cal Q}_z^{2})\label{Hamil1}\\
&=&\frac{1}{2}[\frac{{\cal P}_{x}^{2}+
{\cal P}^{2}_{y}}{m}+K'_{x}{\cal Q}_x^{2}+K'_{y}{\cal Q}_y^{2}-\Omega
{\cal L}_z]+{\cal H}_z\,,\label{H2}
\end{eqnarray}
where $\mathbf{A}=\frac{1}{2}\bm{H}\times\bm{Q}$ is the vector potential,
$\Omega=\frac{e|\mathbf{H}|}{mc}$ the cyclotron frequency, ${\cal L}_z={\cal
Q}_x{\cal P}_y-{\cal Q}_y{\cal P}_x$, ${\cal H}_z= \frac{1}{2}(\frac{{\cal
P}_z^{2}}{m} +K_z {\cal Q}_z^{2})$  and $K'_{\mu}=K_{\mu}+m\Omega^2/4$. Eq.\
(\ref{H2}) is also identical with the intrinsic Hamiltonian of a particle of
mass $m$ in a quadratic potential of constants $K'_\mu$ rotating around the $z$
axis with frequency $\Omega/2$ \cite{RS.80,BR.86}. 

The replacement ${\cal P}_\mu=P_\mu\sqrt{\hbar m\Omega_0}$, ${\cal Q_\mu}
=Q_\mu/\sqrt{m\Omega_0/\hbar}$ in (\ref{H2}), with $\Omega_0$ a reference
frequency, leads to ${\cal H}=\hbar \Omega_0 H+{\cal H}_z$, with $H$ given by
(\ref{H1}) and
 \begin{equation}
 k'_\mu=k_\mu+\omega^2\,,\;\;k_\mu=\frac{K_\mu}{m\Omega_0^2}
 \,,\;\;\omega=\frac{\Omega}{2\Omega_0}\,.\label{kmu}
 \end{equation}

We note that in terms of the boson operators  $b_\mu=(Q_\mu+iP_\mu)/\sqrt{2}$,
the scaled angular momentum $L_z={\cal L}_z/\hbar$ in (\ref{H1}) is
\begin{equation}
L_z=Q_xP_y-Q_yP_x=-i(b^\dagger_x b_y-b^\dagger_y b_x)
\,,\label{Lz}\end{equation}
and can then be simulated by standard linear optics, although for such bosons
the first two terms in (\ref{H1}) become $\sum_{\mu=x,y} g_\mu^+(b^\dagger_\mu
b_\mu+\frac{1}{2})+ g_\mu^-(b_\mu^2+{b^\dagger_\mu}^2)$, with
$g_\mu^\pm=\frac{k'_\mu\pm 1}{2}$, and require non-linear means. If $K'_x>0$,
we can set $k'_x=1$, i.e., $g_x^-=0$, by adequately fixing $\Omega_0$ in
(\ref{kmu}), but $g_y^-$ remains non-zero in the relevant anisotropic case
$k'_y\neq k'_x$, where $[H,L_z]\neq 0$. The change to normal $x,y$ bosons, such
that $H=\sum_{\mu=x,y} \sqrt{k'_\mu}(\tilde{b}^\dagger_\mu
\tilde{b}_\mu+\frac{1}{2})-\omega L_z$, will lead instead to an additional 
term $\propto (1-\sqrt{\frac{k'_x}{k'_y}})(\tilde{b}_x
\tilde{b}_y-\tilde{b}^\dagger_x\tilde{b}^\dagger_y)$ in $L_z$.

\subsection{Diagonalization and stability}
If the parameter
\begin{equation}
\Delta=\sqrt{(k'_x-k'_y)^2/4+2\omega^2(k'_x+k'_y)}\end{equation}
is {\it non-zero}, the canonical transformation
\begin{eqnarray}
P'_{\mu}&=&P_{\mu}+\gamma Q_{-\mu}\,,\;\;\;\; Q'_{\mu}=\frac{Q_{\mu}-\eta
P_{-\mu}}{1+\eta\gamma} \label{qp2}\,,
\end{eqnarray}
where $\gamma=\frac{2\Delta - k'_x+k'_y}{4\omega}$, $\eta =
\frac{2\gamma}{k'_x+k'_y}$ and
labels $(x,y)$ are now identified  with $(+,-)$,
allows to write Eq.\ (\ref{H1}) as  \cite{RK.09}
\begin{eqnarray}
H&=&\sum_{\mu=\pm} \frac{1}{2}(\alpha_\mu {P'}_\mu^2+\beta_\mu
{Q'}_\mu^2)\,,\label{H3}
\end{eqnarray}
where $\alpha_{\pm}=1-{\textstyle\frac{\omega}{\Delta}}(\gamma\mp\omega)$ and
$\beta_{\pm}={\textstyle\frac{\Delta}{\omega}}(\gamma\pm\omega)$. If $\Delta=0$
and $\omega\neq 0$, a separable representation of the form (\ref{H3}) in terms
of canonical variables ($[Q'_\mu,P'_\nu]=\delta_{\mu\nu}$,
$[P'_\mu,P'_\nu]=[Q'_\mu,Q'_\nu]=0$) {\it is not feasible}. Such a possibility
can arise in the repulsive case $k_\mu\leq 0$, when the $4\times 4$ matrix
representing the quadratic form (\ref{H1}) is not diagonalizable with the
symplectic metric and leads to non-trivial Jordan forms  \cite{RK.09}.

For general real values of $k'_\mu$ in (\ref{H1}), the coefficients
$\alpha_{\mu}$, $\beta_{\mu}$ in (\ref{H3}) can be positive, zero, negative,
and even complex \cite{RK.09}.  We will here consider those cases where Eq.\
(\ref{H3}) can be further written as
\begin{eqnarray}
H&=&\sum_{\mu=\pm} \lambda_\mu(b'^\dagger_\mu b'_\mu+\frac{1}{2})\,,\label{H4}\\
|\lambda_{\pm}|&=&\sqrt{\alpha_{\pm}\beta_{\pm}}=
\sqrt{{\textstyle\frac{k'_x+k'_y}{2}}+\omega^2\pm\Delta}\,,\label{lam}
\end{eqnarray}
with $\lambda_{\mu}$ real and
${b'}_\mu=\sqrt\frac{\beta_\mu}{2\lambda_\mu}Q'_\mu+
i\sqrt{\frac{\alpha_\mu}{2\lambda_\mu}}P'_\mu$ standard bosons
($[{b'}_\mu,{b'}_\nu^\dagger]=\delta_{\mu\nu}$,
$[{b'}_\mu,{b'}_\nu]=[{b'}^\dagger_\mu,{b'}^\dagger_\nu]=0$), such that $H$
exhibits a {\it discrete} spectrum. In these cases the matrix representing
(\ref{H1}) is diagonalizable with the symplectic metric, with real symplectic
eigenvalues \cite{RK.09}.

\begin{figure}[t] \centering
\hspace*{0.cm}\includegraphics[trim=0cm 0cm 0cm 0cm, width = 9cm]{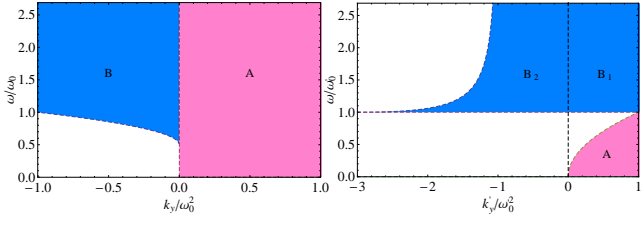}
\vspace*{-.75cm}

\caption{Left: Sectors with discrete spectrum at fixed $k_\mu$
(particle in a magnetic field plus quadratic potential) in the
anisotropy-frequency plane $k_y/|k_x|,\omega/\omega_0$, where
$\omega_0=\sqrt{|k_x|}$. $A$ is the positive definite sector, corresponding to
an attractive potential $k_\mu>0$ for $\mu=x,y$, whereas $B$ the non-positive
sector, corresponding to a repulsive potential $k_\mu<0$ for $\mu=x,y$ with
$|\omega|>\omega_c$ (Eq.\ (\ref{wc})). Right: Same sectors at fixed $k'_\mu>0$
(particle in a rotating quadratic potential) in the
$k'_y/k'_x,\omega/\omega'_0$ plane, where $\omega'_0=\sqrt{k'_x}$ and we have
set $k'_x>0$. $A$ is the positive definite sector (Eq.\ (\ref{wmin})), $B_1$
the non-positive sector with $k'_y>0$ (Eq.\ (\ref{wmax})) and $B_2$ that with
$k'_y<0$ (Eqs.\ (\ref{we0})--(\ref{we})). Quantities plotted in all figures are
dimensionless.} \label{f1}
\end{figure}

At fixed $k_\mu$ in (\ref{kmu}) (charged particle in
a magnetic field), Eq.\ (\ref{H4}) is valid in the following domains
\cite{RK.09} (Fig.\ \ref{f1}):\\
(A) $k_{x}>0$, $k_y>0$, where $\alpha_\pm>0$,  $\beta_\pm>0$ and
$\lambda_\pm>0$. This is the standard case of an attractive quadratic potential,
where $H$ is positive definite and hence fully stable. \\
(B) $k_{x}<0$, $k_y<0$, and
\begin{equation}
|\omega|>\omega_{c}={\textstyle\frac{\sqrt{-k_x}+\sqrt{-k_y}}{2}}\,,
 \label{wc}\end{equation}
where $\alpha_+>0$, $\beta_+>0$ but $\alpha_-<0$, $\beta_-<0$, implying
$\lambda_+>0$ but $\lambda_-<0$ in (\ref{H4}). This is the case of a {\it
repulsive} quadratic potential, where $H$ becomes equivalent to a standard plus
an {\it inverted} oscillator if $|\omega|>\omega_c$: It has no minimum energy,
but is \textit{dynamically stable}, as the motion remains bounded \cite{RK.09}.
The dynamics around a quadratic potential maximum can then be
stabilized by a sufficiently strong field.\\
(C) $k_x=k_y=0$ and $\omega\neq 0$ (\textit{Landau case}), where
$\Delta=2\omega^2$, $\alpha_+=1$, $\beta_+=4\omega^2$ whereas
$\alpha_-=\beta_-=0$,  leading to $\lambda_+=2|\omega|$ and $\lambda_-=0$. Eq.\
(\ref{H4}) becomes a standard plus a vanishing oscillator. Here the final
choice of $b'_-$, ${b'}_-^\dagger$ is not fixed, as $\lambda_-=0$. We will set
$b'_-=\sqrt{\omega}Q'_-+iP'_-/\sqrt{4\omega}$, according to the
$k_\mu\rightarrow 0$ limit of the isotropic case $k_x=k_y=k$, where
$\lambda_{\pm}=\sqrt{k'}\pm\omega$, $\Delta=2\omega\sqrt{k'}$,
$\gamma=\sqrt{k'}=1/\eta$, $\alpha_{\mu}=\frac{1}{2}\lambda_{\mu}/\sqrt{k'}$,
$\beta_{\mu}=2\lambda_{\mu}\sqrt{k'}$ and $k'=k+\omega^2$.

\vspace*{-0.cm}

At fixed $k'_\mu$ in (\ref{H1}) (rotating potential), the previous sectors are
seen quite differently. Sector (A) corresponds to  $k'_x>0$, $k'_y>0$ {\it and}
\begin{equation}
|\omega|<\omega'_{c1}={\rm Min}[\sqrt{k'_x},\sqrt{k'_y}]\,,\label{wmin}
\end{equation}
indicating a {\it maximum allowable frequency} in an attractive rotating quadratic
potential (right panel in Fig.\ \ref{f1}). Sector (B) corresponds to
\begin{eqnarray}
|\omega|&>&\omega'_{c2}={\rm Max}
 [\sqrt{k'_x},\sqrt{k'_y}]\label{wmax}\end{eqnarray}
if $k'_x>0$, $k'_y>0$. Thus, as the frequency $|\omega|$ is increased above
$\omega'_{c1}$ a {\it finite instability interval}
$\omega'_{c1}<|\omega|<\omega'_{c2}$ arises in the anisotropic case $k'_x\neq
k'_y$, although dynamical stability is again recovered for
$|\omega|>\omega'_{c2}$. In addition, sector (B) also corresponds here to
$k'_x>0$ and $k'_y<0$ (or viceversa), provided \cite{RK.09}
\begin{eqnarray}
&|\omega|&>\sqrt{k'_x}\,,\;- k'_x\leq k'_y<0\,,\label{we0}\\
\sqrt{k'_x}&<&|\omega|<\omega'_{c3}=\frac{k'_x-k'_y}{\sqrt{8(k'_x+k'_y)}}\,,\;
 -3k'_x<k'_y<-k'_x.\label{we}\end{eqnarray}
Hence, a quadratic potential repulsive in one of the axes can be stabilized by
increasing the frequency above $\sqrt{k'_x}$ if $-3k'_x<k'_y<0$, although
stability holds just within a finite interval if $-3k'_x<k'_y<-k'_x$. Finally,
the Landau case (C) corresponds to $k'_x=k'_y=\omega^2$.

\subsection{Covariance matrix}
Both the vacuum $|0'\rangle$  of the primed bosons $b'_\mu$ in (\ref{H4}), and
the thermal state
\begin{eqnarray}
 \rho&=&Z^{-1}\exp[-\beta H]\,,\label{T}
 \end{eqnarray}
well defined in the stable region (A) (with $\beta=1/T>0$ and $Z={\rm
Tr}\,\exp[-\beta H]=\sum_{\mu=\pm}\frac{1}{2\sinh(\beta\lambda_\mu/2)}$), are
{\it gaussian} states \cite{RS.00,WW.01,AEPW.02,ASI.04,CEPD.06,AI.08,MRC.10}.
Any expectation value, and in particular the entanglement between the $x$ and
$y$ modes in these states, will then be completely determined by the elements
of the basic covariance matrix of the operators $Q_\mu$, $P_\mu$, which we
define as \cite{MRC.10} (note that here $\langle Q_\mu\rangle=\langle
P_\mu\rangle=0$)
\begin{eqnarray}
{\cal D}^R&=&\langle RR^t\rangle-{\cal M}^R= \left(\begin{array}{cc}
Q&L\\L^t&P\end{array}\right) -{\textstyle\frac{1}{2}}{\cal M}^R\,,
 \label{DR}\\
 {\cal M}^R&=&{R}{R}^t-(RR^t)^t=
i\left(\begin{array}{cc}0&1\\-1&0\end{array}\right)\,,
\label{MR}\end{eqnarray}
where $R=(Q_x,Q_y,P_x,P_y)^t$ and hence $Q_{\mu\nu}=\langle Q_\mu
Q_\nu\rangle$, $P_{\mu\nu}=\langle P_\mu\,P_\nu\rangle$ and  $L_{\mu\nu}=\langle
Q_\mu P_\nu+P_\nu Q_\mu\rangle/2$.

Eq.\ (\ref{DR}) is unitarily related to the non-negative bosonic contraction
matrix \cite{RS.80,MRC.10}
\begin{eqnarray}
{\cal D}&=&\langle Z Z^\dagger\rangle-{\cal M}
=\left(\begin{array}{cc} F&G\\\bar{G}&1+\bar{F}\end{array}\right)\,,\\
{\cal M}&=&{Z}{Z}^\dagger-((Z^\dagger)^t Z^t)^t=
 \left(\begin{array}{cc}1&0\\0&-1\end{array}\right)\,,\end{eqnarray}
where $Z=(b_x,b_y,b^\dagger_x,b^\dagger_y)^t$, $F_{\mu\nu}=\langle
b^\dagger_\nu b_\mu\rangle$,  $G_{\mu\nu}=\langle b_\mu b_\nu\rangle$ and
$b_\mu=(Q_\mu+i P_\mu)/\sqrt{2}$. Since $R={\cal U}Z$, with ${\cal
U}=(^{\;\;1\;1}_{-i\;i})/\sqrt{2}$, we have ${\cal D}^R={\cal U}{\cal D}{\cal
U}^\dagger$  and ${\cal M}^R={\cal U}{\cal M}{\cal U}^\dagger$.

In both the vacuum and the thermal state (\ref{T}), we have
\begin{equation}
\langle b'_\mu b'_\nu\rangle=0\,,\;\;
\langle {b'}^\dagger_\mu b'_\nu\rangle=f'_\mu\delta_{\mu\nu}\,,
\label{f}\end{equation}
where $f'_\mu=0$ in the vacuum state and
\begin{equation}
 f'_\mu=-\frac{1}{\beta}\frac{\partial\ln Z}{\partial\lambda_\mu}
 -\frac{1}{2}=\frac{1}{e^{\beta\lambda_\mu}-1}\,,
 \label{ft}\end{equation}
in the thermal state. By inverting Eq.\ (\ref{qp2}), we then obtain $\langle
Q_{x}Q_{y}\rangle=\langle P_x P_y\rangle=L_{\mu\mu}=0$ for $\mu=x,y$, and
\begin{eqnarray}
\langle Q_{\mu}^2\rangle&=&
\langle {Q'}_{\mu}^2\rangle+\frac{\eta^2}{(1+\gamma\eta)^2}
\langle {P'}_{-\mu}^2\rangle\,,\label{Qu}\\
\langle P_{\mu}^2\rangle&=&
\frac{1}{(1+\gamma\eta)^2}\langle {P'}_{\mu}^2\rangle
+\gamma^2\langle {Q'}_{-\mu}^2\rangle\,,\\
 \langle Q_{\mu}P_{-\mu}\rangle&=&
 -\gamma\langle {Q'}_{\mu}^2\rangle
 +\frac{\eta}{(1+\gamma\eta)^2}\langle {P'}_{-\mu}^2\rangle\,,
 \label{qe}\end{eqnarray}
where
 \begin{eqnarray}
 \langle {Q'}_{\mu}^2\rangle&=& (f'_{\mu}+{\textstyle\frac{1}{2}})
 \frac{\lambda_{\mu}}{\beta_{\mu}},\;\; \langle { P'}_{\mu}^2\rangle=(f'_{\mu}+
 {\textstyle\frac{1}{2}})\frac{\lambda_\mu}{\alpha_\mu}.
 \label{qep}\end{eqnarray}
These averages provide all the elements of (\ref{DR}). The symplectic
eigenvalues of  ${\cal D}^R$ and ${\cal D}$ are coincident and given precisely
by $f'_\mu$ and $-1-f'_\mu$ (Eqs.\ (\ref{f})--(\ref{ft})), with physical states
corresponding to $f'_\mu\geq 0$. They are just the standard eigenvalues of the
matrix ${\cal D}{\cal M}=(^{F\;\;-G}_{\bar{G}\;-1-\bar{F}})$,
 or equivalently, ${\cal D}^R{\cal M}^R={\cal U}{\cal
D}{\cal M}{\cal U}^\dagger=i(^{-L\;Q}_{-P\;L^t})-I/2$.

\subsection{Vacuum entanglement}
The entanglement of the vacuum $|0'\rangle$ is a measure of its deviation from
a product state  $|0_x\rangle\otimes |0_y\rangle$. It can be quantified through
the entanglement entropy \cite{BBPS.96}, which is just the von Neumann entropy
of the reduced state $\rho_\mu={\rm Tr}_{-\mu}\,\rho$ of any of the modes
($\mu=x,y$), since for a pure state ($\rho=|0'\rangle\langle 0'|$) they are
isospectral. The state $\rho_\mu$ is a gaussian mixed state completely
determined by the reduced $2\times 2$ covariance matrix
\begin{eqnarray}
{\cal D}^R_\mu&=&\left(\begin{array}{cc} \langle Q_\mu^2\rangle &L_{\mu\mu}\\
L_{\mu\mu}&\langle
P_\mu^2\rangle\end{array}\right)-{\textstyle\frac{1}{2}}{\cal M}^R\,,
 \label{DRr} \end{eqnarray}
whose symplectic eigenvalues are $f_\mu={\rm det}^{1/2}[{\cal
D}^R_\mu+\frac{1}{2}{\cal M}^R]-\frac{1}{2}$ and $-1-f_\mu$. Here
$L_{\mu\mu}=0$ and hence,
\begin{eqnarray}
f_\mu&=&\sqrt{\langle Q_\mu^2\rangle
\langle P_\mu^2\rangle}-{\textstyle\frac{1}{2}}\,,\label{fmu}
\end{eqnarray}
which is just the deviation of the mode uncertainty from its minimum value. The
entropy of $\rho_\mu$ is, therefore, that of a boson system with average
occupation $f_\mu$:
\begin{eqnarray}
S(\rho_\mu)&=&-{\rm Tr}\,\rho_\mu\log\rho_\mu=h(f_\mu)\,,\label{Smu}\\
h(f)&=&-f\log f+(1+f)\log (1+f)\,,\label{h}
\end{eqnarray}
which is just a positive concave increasing function of $f_\mu$. The vacuum is
then entangled iff $f_\mu>0$, with $S\approx -f_\mu(\log f_\mu-1)$ for
$f_\mu\rightarrow 0$ and $S\approx \log f_\mu+1$ for $f_\mu\rightarrow\infty$
(for $\log=\ln$).

In the vacuum case ($f'_\mu=0$) Eqs.\ (\ref{Qu})--(\ref{fmu}) lead to
\begin{equation}
f_\mu=\frac{1}{2}\left[\frac{\bar{\omega}}{\bar{\omega}_g}
\sqrt{\frac{\bar{\omega}_g^2+\omega^2}
{\bar{\omega}^2+\omega^2}}-1\right]
 \,,\label{fmu3}\end{equation}
which is independent of $\mu$, where
\begin{equation}
\bar{\omega}={\textstyle\frac{\omega_x+\omega_y}{2}},~~~~
\bar{\omega}_g=\sqrt{\omega_x\omega_y}\,,\label{wav}
\end{equation}
denote the arithmetic and geometric averages of the original oscillator
frequencies $\omega_\mu=\sqrt{k_\mu}$. Entanglement is thus completely
determined by the ratios $\omega/\bar{\omega}$ and
$\bar{\omega}_g/\bar{\omega}$ (with $\bar{\omega}_g/\bar{\omega}\leq 1$), or
equivalently, $\omega/\omega_x$ and $\omega_y/\omega_x$. It is then {\it
non-zero $\forall$ $\omega>0$ if $\omega_x\neq\omega_y$}, i.e.,
$\bar{\omega}\neq \bar{\omega}_g$ (anisotropic case).  In sector $A$, the
$\omega_\mu$ are positive, whereas in $B$ they are both imaginary, implying
\begin{equation}
f_\mu=\frac{1}{2}\left[\frac{|\bar{\omega}|}{|\bar{\omega}_g|}
\sqrt{\frac{\omega^2-|\bar{\omega}_g|^2}
{\omega^2-|\bar{\omega}|^2}}-1\right]\;\;\;(k_\mu<0)\,.
\end{equation}
\subsection{Thermal entanglement}
For a mixed bipartite state, like the thermal state (\ref{T}) at $T>0$,
entanglement is a measure of its deviation from a {\it separable} state
\cite{RW.89}, i.e., from a convex combination of product states
$\rho_s=\sum_\alpha q_\alpha \rho_x^\alpha\otimes\rho_y^\alpha$, where
$q_\alpha>0$,  $\sum_\alpha q_\alpha=1$. Such states can be created by local
operations and classical communication. For a two-mode gaussian mixed state,
entanglement can be quantified by the {\it negativity} \cite{VW.02}, which is
minus the sum of the negative eigenvalues of the partial transpose $\rho^{t_y}$
of the total density matrix $\rho$, measuring then the degree of violation of
the PPT criterion \cite{PP.96,HHH.96} by the entangled state. For a two-mode
gaussian state, a positive negativity is a necessary and sufficient condition
for entanglement \cite{RS.00}.

Partial transposition with respect to $y$ implies the replacement
$P_y\rightarrow-P_y$ in the full covariance matrix (\ref{DR})
\cite{RS.00,WW.01,AEPW.02,ASI.04}, leading to a matrix $\tilde{\cal D}^R$. The
negativity can then be evaluated in terms of the negative symplectic
eigenvalues of this  matrix, which will have eigenvalues $\tilde{f}_\mu$ and
$-1-\tilde{f}_\mu$, with $\tilde{f}_\mu\geq -1/2$ \cite{MRC.10}. Replacing
$L_{xy}$ by $-L_{xy}$ in (\ref{DR}), we obtain here
\begin{eqnarray}
\tilde{f}_\pm&=&\sqrt{\tilde{\alpha}\pm
\sqrt{\tilde{\alpha}^2-\beta^2}}-{\textstyle\frac{1}{2}}\nonumber\\
&=&{\textstyle\sqrt{\frac{1}{2}(\tilde{\alpha}+\beta)}\pm
\sqrt{\frac{1}{2}(\tilde{\alpha}-\beta)}-\frac{1}{2}}\,,\label{fti}\end{eqnarray}
where $\tilde{\alpha}$ and $\beta$ can be expressed in terms of the local
symplectic eigenvalues $f_\mu$ (\ref{fmu}) and the global ones $f'_\mu$ (\ref{ft}):
\begin{eqnarray}
\tilde{\alpha}&=&{\textstyle\frac{1}{2}}(\langle Q_x^2\rangle\langle
P_x^2\rangle+
\langle Q_y^2\rangle\langle P_y^2\rangle)+
\langle Q_x P_y\rangle\langle Q_y P_x\rangle\nonumber\\
&=&{\textstyle\sum_\mu[(f_\mu+\frac{1}{2})^2-\frac{1}{2}(f'_\mu+\frac{1}{2})^2]}
\,,\label{alp}\\
\beta&=&\sqrt{(\langle Q_x P_y\rangle^2-\langle Q_x^2\rangle\langle P_y^2\rangle)
(\langle Q_y P_x\rangle^2-\langle Q_y^2\rangle\langle P_x^2\rangle)}\nonumber\\
&=&{\textstyle\prod_\mu(f'_\mu+\frac{1}{2})}\,.\label{bef}
\end{eqnarray}
(Note that if $\tilde{\alpha}$ is replaced by
$\alpha={\textstyle\frac{1}{2}}(\langle Q_x^2\rangle\langle
P_x^2\rangle+\langle Q_y^2\rangle\langle P_y^2\rangle)- \langle Q_x
P_y\rangle\langle Q_y P_x\rangle$, Eq.\ (\ref{fti}) becomes $f'_{\pm}$). While
$f_\mu$ depends on $\mu$ for $T>0$, Eq.\ (\ref{alp}) depends  just on the sum
\begin{eqnarray}&&{\textstyle\sum_\mu (f_\mu+{\textstyle\frac{1}{2}})^2}=
{\textstyle\frac{\bar{\omega}^2(\omega^2+\bar{\omega}_g^2)}
{2\bar{\omega}_g^2(\omega^2+\bar{\omega}^2)}
(1+2\sum_\mu f'_\mu)}+\nonumber\\
&&{\textstyle\frac{(\omega^2-\bar{\omega}_g^2)
(\omega^2+2\bar{\omega}^2)+2\bar{\omega}^4]}
{2(\omega^2+\bar{\omega}^2)(\omega^2+\bar{\omega}^2-\bar{\omega}_g^2)}
 \sum_\mu{f'_\mu}^2}+
{\textstyle\frac{\omega^2[(\omega^2-\bar{\omega}_g^2)
(2\bar{\omega}^2-\bar{\omega}_g^2)
+2\bar{\omega}^4]}
{\bar{\omega}_g^2
(\omega^2+\bar{\omega}^2)(\omega^2+\bar{\omega}^2-\bar{\omega}_g^2)}
\prod_\mu f'_\mu}\,.\nonumber
\end{eqnarray}

The negativity can then be expressed as \cite{MRC.10}
\begin{eqnarray}
N&=&\frac{1}{2}({\rm Tr}|\rho^{t_y}|-1)=
\frac{1}{2}[\prod_{\mu}\frac{1}{1+\tilde{f}_\mu-|\tilde{f}_\mu|}-1]
\nonumber\\
 &=&{\rm Max}[\frac{-\tilde{f}_-}{1+2\tilde{f}_-},0]
 \,,\label{Nt}\end{eqnarray}
since only $\tilde{f}_-$ can be negative. The entanglement condition
$\tilde{f}_-<0$ leads to $\tilde{\alpha}>\frac{1}{8}+2\beta^2$ or
\begin{equation}
\sum_\mu f_\mu(1+f_\mu)>\sum_\mu f'_\mu(1+f'_\mu)+2\prod_\mu
f'_\mu(1+f'_\mu)\,,\label{TE}
\end{equation}
which imposes a temperature dependent lower bound on the average local
occupation.

In the vacuum case  $f'_\mu=0$,  Eq.\ (\ref{TE}) implies just $f_\mu>0$,
while  Eqs.\ (\ref{fti})--(\ref{Nt})  reduce to
\begin{equation}
 \tilde{f}_-=f_\mu-\sqrt{f_\mu(f_\mu+1)}\,,\;\;
 N=f_\mu+\sqrt{f_\mu(f_\mu+1)}\,,
 \label{ftvac}\end{equation}
with $f_\mu$ given by Eq.\ (\ref{fmu3}), in agreement with the general results
for pure gaussian states \cite{MRC.10}. Both $-\tilde{f}_-$ and $N$ are again
concave increasing functions of $f_\mu$ at $T=0$ and can be taken as alternative
vacuum entanglement measures.

\subsection{Quantum Discord\label{QDs}}

Quantum discord \cite{OZ.01} is essentially a measure of the deviation of a
bipartite mixed quantum state from a {\it classically correlated state}, i.e.,
a state diagonal in a standard or conditional product basis.  For a general
bipartite system $A+B$, the quantum discord $D^B$ can be defined as the minimum
difference between the conditional von Neumann entropy of $A$ after an unread
local measurement $M_B$ in $B$ and the original quantum conditional entropy
$S(A|B)=S(A,B)-S(B)$ \cite{OZ.01}:
\begin{equation}
D^B=\mathop{\rm Min}_{M_B}\sum_j p_j S(\rho_{A/j})-[S(\rho_{AB})-S(\rho_B)]\,,
\label{DB}
\end{equation}
where, for a measurement $M_B$ based on local projectors $P_j$ ($\sum_j P_j=
I_B$), $p_j={\rm Tr}\rho_{AB} I_A\otimes P_j$ is the probability of outcome $j$
and $\rho_{A/j}$ the reduced state of $A$ after such outcome. Eq.\ (\ref{DB})
can be also expressed as the minimum difference between the original mutual
information $I(A:B)=S(A)-S(A|B)$, which measures all correlations between $A$
and $B$, and that after the unread local measurement, $S(A)-\sum_j p_j
S(\rho_{A/j})$, which contains the ``classical'' part of the quantum
correlations \cite{OZ.01,HV.01}.

For a pure state ($\rho_{AB}^2=\rho_{AB}$) both $S(\rho_{AB})$ and
$S(\rho_{A/j})$ vanish and $D^B$ reduces to the entanglement entropy
$S(\rho_B)=S(\rho_A)$, with $D^A=D^B$ \cite{OZ.01}. For a mixed state, however,
$D^B$ is not an entanglement measure, being in fact non-zero for most separable
states \cite{FA.10} and vanishing just for those separable states of the form
$\rho_c=\sum_j p_j\rho_{A/j}\otimes P_j$ (classically correlated with respect
to $B$), which remain unaltered after the local measurement $M_B$. In general,
$D^B\neq D^A$ for mixed states. Hence, for a bipartite system with a
non-degenerate ground state in a thermal equilibrium state, like the system
under study, differences between quantum discord and entanglement, and between
$D^A$ and $D^B$, will arise only at finite temperature.

The exact evaluation of $D^B$ involves a difficult minimization over all local
measurements $M_B$. Nonetheless, for a two-mode gaussian state, a minimization
restricted to gaussian measurements was recently shown to be analytically
feasible \cite{GP.10,AD.10,AD.11}. For such measurements in the present system,
Eq.\ (\ref{DB}) becomes, choosing $B=y$ and using Eqs.\ (\ref{ft}),
(\ref{Smu})--(\ref{h}),
\begin{equation}
D^y={\rm Min}_{M_y} h(f_{x}^{M_y})-[h(f'_+)+h(f'_-)-h(f_y)], \label{Dy}
\end{equation}
where $f_x^{M_y}$ denotes the symplectic eigenvalue of the covariance matrix
${\cal D}^{M_y}_x$ associated with $\rho_{x/j}$, which  depends on the $2\times
2$ covariance matrix ${\cal D}_{M_y}$ determining the local gaussian
measurement $M_y$ \cite{GP.10,AD.10}. The final result was provided in
\cite{AD.10} and can be fully expressed in terms of the local invariants
$A=4(f_x+\frac{1}{2})^2$, $B=4(f_y+\frac{1}{2})^2$, $C= 2\sum_\mu
(f'_\mu+\frac{1}{2})^2-(f_\mu+\frac{1}{2})^2$ and $D=\prod_\mu
4(f'_\mu+\frac{1}{2})^2$, which determine the quantity $E_{\rm Min}$
of \cite{AD.10},  with ${\rm Min}_{M_y} f_x^{M_y}=\frac{1}{2}\sqrt{E_{\rm
Min}}-\frac{1}{2}$. It can be shown that if  $D^y> 1$ the two-mode gaussian
state is entangled \cite{GP.10,AD.10}. Moreover, the only two-mode gaussian
states with  $D^y=0$ are product states  \cite{AD.10}. The expression for $D^x$
(local measurement in $x$) is obviously similar ($x\leftrightarrow y$ in
previous formulas).

\section{Results \label{III}}
\subsection{Vacuum entanglement}
Let us now analyze the main features of Eq.\ (\ref{fmu3}). We first consider
{\it fixed $k_\mu$} in (\ref{kmu}) (charged particle in a magnetic field). In
the isotropic case $\omega_x=\omega_y$, $\bar{\omega}=\bar{\omega}_g$ and
$f_\mu=0$ $\forall$ $\omega$. There is no entanglement since $L_z$ commutes in
this case with $H$ and leaves the isotropic product vacuum invariant. For
$|\omega_x-\omega_y|\ll \bar{\omega}$, Eq.\ (\ref{fmu3}) leads to
\begin{equation}
f_\mu\approx{\frac{\omega^2}{16\bar{\omega}^2(\bar{\omega}^2+\omega^2)}}
(\omega_x-\omega_y)^2+O((\omega_x-\omega_y)^4)\,,
\end{equation}
indicating a quadratic vanishing of $f_\mu$ in this limit. Entanglement also
vanishes for $\omega\rightarrow 0$ (no coupling), where
\begin{equation}
{f_\mu\approx\frac{1}{4}
~\left(\frac{1}{\bar{\omega}_g^2}-\frac{1}{\bar{\omega}^2}\right)\omega^2+O(\omega^4)
 \,.\label{f0} }\end{equation}

\begin{figure}[t] \centering
\includegraphics[trim=0cm 0cm 0cm 0cm, width = 7cm]{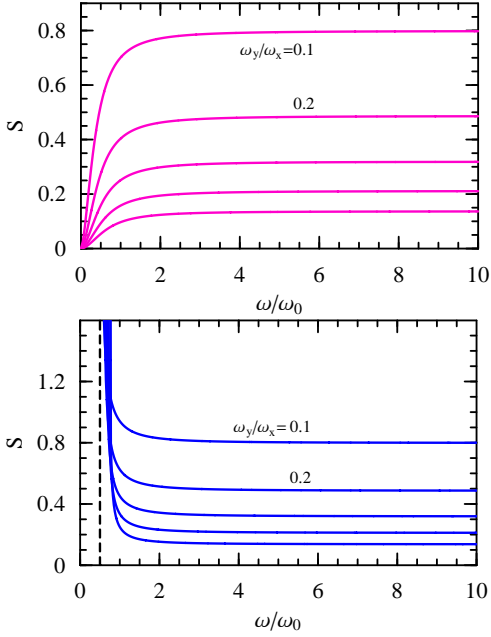}
\vspace*{-.75cm}

\caption{Entanglement entropy $S=S(\rho_\mu)$ (Eq.\ (\ref{Smu}))
between the two modes as a function of $\omega$ in the vacuum of Hamiltonian
(\ref{H1}) at fixed $k_\mu=\omega_\mu^2$, for  ratios
$\omega_y/\omega_x=0.1,\ldots,0.5$ and $\omega_0=|\omega_x|$.  The top panel
corresponds to sector A (attractive potential $k_\mu>0$), the bottom panel to
sector B (repulsive potential $k_\mu<0$, $|\omega|>\omega_c$), where $S$
diverges for $\omega\rightarrow \omega_c$ (Eq.\ (\ref{wc})). In both cases $S$
approaches the same finite limit for $\omega\rightarrow\infty$.} \label{f2}
\end{figure}

On the other hand, for $\omega\rightarrow\infty$, a remarkable feature is that
$f_\mu$ approaches a {\it finite limit}, which depends just on the anisotropy
$\omega_y/\omega_x$: For $\omega\gg |\bar{\omega}|$ Eq.\ (\ref{fmu3}) leads to
\begin{eqnarray}
{f_\mu\approx\frac{1}{2}\left[\frac{\bar{\omega}}{\bar{\omega}_g}-1\right]
 +O(\omega^{-2})}\,.\label{finf}\end{eqnarray}
In sector $A$, $f_\mu$ and hence $S(\rho_\mu)$ are then  {\it increasing}
functions of $\omega$ (Fig.\ \ref{f2}). Mode entanglement is then enhanced just
by increasing the field, although it will {\it saturate} for strong fields.
This saturation is a consequence of the balance between the oscillator part and
the coupling $\omega L_z$ in (\ref{H1}), as $k'_\mu=k_\mu+\omega^2$ also
becomes large, reducing $\langle Q_\mu^2\rangle$: For
$\omega\rightarrow\infty$, $\langle Q_\mu^2\rangle\approx
\frac{\bar{\omega}}{2\omega_{\mu}}\omega^{-1}\rightarrow 0$ while $\langle
P_\mu^2\rangle\approx
\frac{\bar{\omega}}{2\omega_{-\mu}}\omega\rightarrow\infty$, leading to the
finite limit (\ref{finf}).

In contrast, $f_\mu$, and hence  entanglement, will {\it diverge} at the edges
of the dynamically stable region. For instance, if $\omega_y\rightarrow 0$,
$\bar{\omega}\rightarrow\omega_x/2$ whereas $\bar{\omega}_g\rightarrow 0$,
implying $f_\mu\propto 1/\sqrt{\omega_y}$:
\begin{equation}
{f_\mu\approx
\frac{1}{2}\left[\sqrt{\frac{\bar{\omega}}{2\omega_y}}\frac{\omega}{
\sqrt{\bar{\omega}^2+\omega^2}}-1\right]
 \,, }\end{equation}
and hence $S(\rho_\mu)\approx \frac{1}{2}\log(\omega_x/\omega_y)$ plus constant
terms. This divergence stems from that of $\langle Q_y^2\rangle$ (or $\langle
P_x^2\rangle$) in this limit, with $\langle P_y^2\rangle$ and $\langle
Q_x^2\rangle$ remaining constant (Eqs.\ (\ref{Qu})--(\ref{qep})).

In the repulsive sector $B$, $f_\mu$ diverges for
$\omega\rightarrow |\bar{\omega}|=\omega_c$ (Eq.\ (\ref{wc})), where {\it both}
$\langle Q_\mu^2\rangle$ and $\langle P_\mu^2\rangle$ diverge:
\begin{equation}
{f_\mu\approx\frac{1}{2}\left[\sqrt{\frac{|\bar{\omega}|}
{\omega-|\bar{\omega}|}}\sqrt{\frac{|\bar{\omega}|^2
-|\bar{\omega}_g|^2}{2|\bar{\omega}_g|^2}}-1\right]
 }\,.\end{equation}
It is then seen that here $f_\mu$ and hence $S(\rho_\mu)$ {\it decrease} as
$\omega$ increases from $\omega_c$ (Fig.\ \ref{f2}, bottom panel), i.e., as the
system becomes dynamically stabilized by the field, reaching for
$\omega\rightarrow\infty$ the same previous limit (\ref{finf}). At  fixed
 $|k_y/k_x|$, the vacuum entanglement is then strictly larger in the
unstable sector $B$ ($k_\mu<0$).

\begin{figure}[t] \centering
\includegraphics[trim=0cm 0cm 0cm 0cm, width = 7cm]{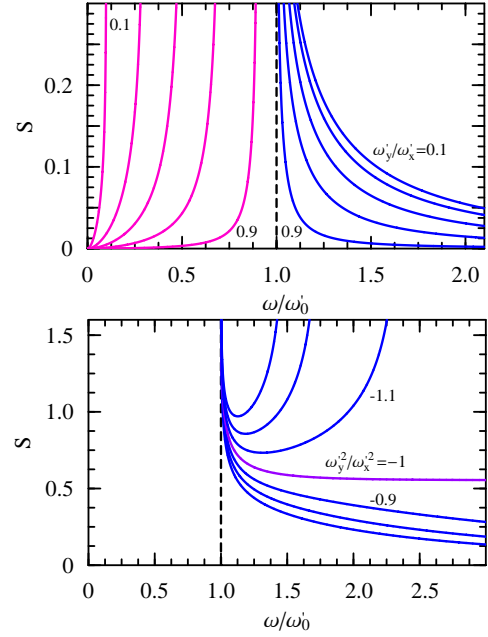}
\vspace*{-.75cm}

\caption{Entanglement entropy between the two modes as a
function of $\omega$ in the vacuum of Hamiltonian (\ref{H1}) at fixed
$k'_\mu={\omega'_\mu}^2$, for ratios $\omega'_y/\omega'_x=0.1,0.3,\ldots,0.9$
(top) and $(\omega'_y/\omega'_x)^2=-1.3,-1.2,\ldots,-0.7$ (bottom), with
$\omega'_0=\omega'_x$. The top panel corresponds to $k'_\mu>0$, with the
positive definite sector A on the left, where $S$ diverges for
$\omega\rightarrow\omega'_{c1}=\omega'_y$, and the non-positive sector $B_1$ on
the right ($\omega>\omega'_{c2}=\omega'_x$), where $S$ diverges at $\omega'_x$
and vanishes for $\omega\rightarrow\infty$. The bottom panel corresponds to
$k'_x>0$ and $k'_y<0$ (sector $B_2$), where $S$ diverges for
$\omega\rightarrow\omega'_x$ and also $\omega\rightarrow\omega'_{c3}$ (Eq.\
(\ref{we})) if $(\omega'_y/\omega'_x)^2<-1$. In the {\it critical case}
$(\omega'_y/\omega'_x)^2=-1$ ($k'_y=-k'_x$), $f_\mu$ and $S$ saturate for large
$\omega$ (Eq.\ (\ref{fcrit})).} \label{f3}
\end{figure}

At {\it fixed $k'_\mu$} (rotating potential) the behavior with frequency is
quite different (Fig.\ \ref{f3}). We should now replace
\begin{equation}
\omega_\mu=\sqrt{{\omega'_\mu}^2-\omega^2},\;\;\omega'_\mu=\sqrt{k'_\mu}
 \,,\end{equation}
in Eqs.\ (\ref{fmu3})--(\ref{wav}). For $\omega'_x=\omega'_y$ there is of
course no entanglement. For $|\omega'_x-\omega'_y|\ll \omega_x$, we have
\begin{equation}
{f_\mu\approx
\frac{\omega^2}{16(\omega^2-{\omega'}_x^2)^2}
(\omega'_x-\omega'_y)^2+O((\omega'_x-\omega'_y)^4)\,.
 }\end{equation}
Entanglement also vanishes for $\omega\rightarrow 0$, where Eq.\ (\ref{f0})
still holds ($\omega'_\mu=\omega_\mu$ at $\omega=0$).

On the other hand, as $\omega$ increases, $f_\mu$ increases rapidly and in
contrast with the previous case, it  {\it diverges} for
$\omega\rightarrow \omega'_{c1}$ (Eq.\ (\ref{wmin})), where, assuming
$\omega'_{c1}=\omega'_y<\omega'_x$,
\begin{equation}
{ f_\mu\approx\frac{1}{2}\left[\sqrt[4]{\frac{{\omega'}^3_y
({\omega'}^2_x-{\omega'}^2_y)}{2(\omega'_y-\omega)
(3{\omega'}_y^2+{\omega'}_x^2)^2}}-1\right]\,, }
 \label{fpc}\end{equation}
implying $S(\rho_\mu)\approx \frac{1}{4}\ln[\omega'_y/(\omega'_y-\omega)]$ plus
constant terms. In this limit $\langle Q_y^2\rangle$ and $\langle P_x^2\rangle$
diverge while $\langle Q_x^2\rangle$ and $\langle P_y^2\rangle$ stay constant,
as $\omega_y=\sqrt{{\omega'_y}^2-\omega^2}\rightarrow 0$.
As $\omega$ increases further, the system enters the instability window,
although for $\omega>{\omega'}_{c2}$ (Eq.\ (\ref{wmax})), it recovers a
discrete spectrum, entering sector $B_1$. For
$\omega\rightarrow{\omega}'_{c2}$, $f_\mu$ diverges  as in
(\ref{fpc}), with $\omega'_y\leftrightarrow\omega'_x$ if
${\omega}'_{c2}=\omega'_x$.

In sector $B_1$, $f_\mu$ and hence the entanglement {\it decrease} as $\omega$
increases, {\it vanishing} for $\omega\rightarrow\infty$, in contrast with the
behavior at fixed $k_\mu$ in sector $A$.  In this limit the vacuum of $H$
becomes now that associated with $\omega L_z$, which is an isotropic
product gaussian state with $L_z=0$, and hence zero entanglement. $\langle
Q_\mu^2\rangle$ and $\langle P_\mu^2\rangle$ stay then finite and their product
approaches minimum uncertainty, leading to
 \begin{equation}
{f_\mu\approx\frac{({\omega'}_x^2-{\omega'}_y^2)^2}
{32\,\omega^2({\omega'}^2_x+{\omega'}_y^2)}+O(\omega^{-4}) }\,.
 \label{fp0} \end{equation}
In the unstable domain $B_2$, the behavior with $\omega$ is the same as in
$B_1$ when $k'_x>0$ and $-k'_x<k'_y<0$.  However, for $k'_x>0$ and $-3k'_x<
k'_y< -k'_x$, we also have the upper instability limit (\ref{we})
($\omega'_{c3}$). In this case $f_\mu$  first decreases with increasing
$\omega$, reaching a {\it minimum}, but then starts again to {\it
increase}, diverging for $\omega\rightarrow\omega'_{c3}$ where now {\it both}
$\langle Q_\mu^2\rangle$ and $\langle P_\mu^2\rangle$ diverge, leading to
\begin{equation}
{f_\mu\approx\frac{1}{2}
\left[\sqrt{\frac{\omega'_{c3}}{2(\omega'_{c3}-\omega)}}-1\right]
 }\,.\label{fp3}\end{equation}
We then obtain {\it different} $O(\omega-\omega_c)^{-1/4}$ and
$O(\omega_c-\omega)^{-1/2}$ divergences of $f_\mu$ at the stability borders
$\omega'_{x}$ and $\omega'_{c3}$ respectively.

In the special {\it critical case} $k'_y=-k'_x$ $(\omega'_y=i\omega'_x$),
where $\omega'_{c3}\rightarrow\infty$, Eq.\ (\ref{fmu3}) leads to
\begin{equation}
{f_\mu=\frac{1}{2}[\sqrt{1+\frac{\omega^2}{\sqrt{\omega^4-{\omega'}_x^2}}}-1]
 }\label{fcrit}\,,\end{equation}
and hence to a {\it finite} asymptotic limit $f_\mu=\frac{1}{2}(\sqrt{2}-1)$
for $\omega\rightarrow\infty$, in contrast with (\ref{fp0}), as also
appreciated in Fig.\ \ref{f3}. In this limit $\langle Q_\mu^2\rangle$ diverges
whereas $\langle P_\mu^2\rangle$ vanishes, the product approaching $1/2$.
Hence, as $\omega$ increases, $f_\mu$ vanishes if $k'_y<-k'_x$,  saturates if
$k'_y=-k'_x$, and diverges (at $\omega=\omega'_{c3}$) if $-3k'_x<k'_y<-k'_x$.

\begin{figure}[t] \centering
\includegraphics[trim=0cm 0cm 0cm 0cm, width = 7cm]{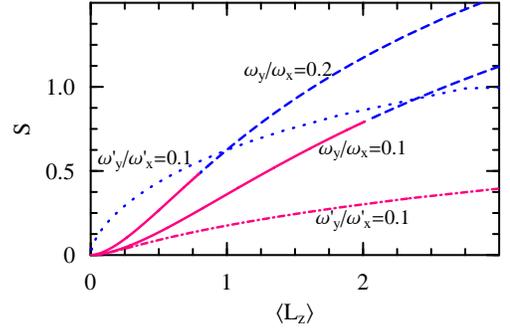}
\vspace*{-.25cm}

\caption{The entanglement entropy between the two modes as a
function of the scaled average angular momentum $\langle L_z\rangle=\langle
{\cal L}_z\rangle/\hbar$ (Eq.\ (\ref{Lz})), in the vacuum of (\ref{H1}) at
fixed $k_\mu=\omega_\mu^2$ (solid and dashed lines) and at fixed
$k'_\mu={\omega'_\mu}^2$ (dashed-dotted and dotted lines), for selected ratios
$\omega_y/\omega_x$ and $\omega'_y/\omega'_x$ and variable $\omega$. The solid
and dashed-dotted lines (in red) correspond to sector A ($k_\mu>0$ or
$k'_\mu>0$  and $|\omega|<\omega'_y$), the dashed and dotted lines (in blue) to
sector B ($k_\mu<0$, $|\omega|>\omega_c$) and $B_1$ ($k'_\mu>0$,
$|\omega|>\omega'_x$). $S$ is in all sectors an increasing function of
$|\langle L_z\rangle|$.} \label{f4}
\end{figure}

The behavior of $f_\mu$ (and hence $S(\rho_\mu)$) with $\omega$ is
qualitatively similar to that of the average angular momentum $\langle
L_z\rangle$. At fixed $k_\mu$, the latter also saturates for
$\omega\rightarrow\infty$ ($\langle L_z\rangle\rightarrow
(\bar{\omega}/\bar{\omega}_g)^2-1$) and diverges for
$\omega\rightarrow\omega_c$ ($\langle L_z\rangle\propto
(\omega-\omega_c)^{-1/2}$),  whereas at fixed $k'_\mu$ it diverges for
$\omega\rightarrow\omega'_{ci}$ ($\langle L_z\rangle\propto
(\omega-\omega'_{ci})^{-1/2}$ for $i=1,2$) and vanishes for
$\omega\rightarrow\infty$. Entanglement is then an {\it increasing} function of
$|\langle L_z\rangle|$ at fixed $k_\mu$ or $k'_\mu$, as seen in Fig.\ \ref{f4},
although it is not fully determined by $|\langle L_z\rangle|$, as the latter is
not invariant under local transformations (in contrast with $f_\mu$). At fixed
$\langle L_z\rangle$, higher ratios $k_y/k_x<1$ originate a higher entanglement
(Fig.\ \ref{f4}). For small $\omega$, $\langle L_z\rangle\propto \omega$ and
hence, $f_\mu\propto \langle L_z\rangle^2$ for small $\langle L_z\rangle$ in
sector $A$. However,  at fixed $k'_\mu$, $\langle L_z\rangle$ also vanishes for
large $\omega$, where $\langle L_z\rangle\propto \omega^{-3}$. Hence, in sector
$B_1$ and according to Eq.\ (\ref{fp0}), $f_\mu\propto\langle L_z\rangle^{2/3}$
for small $\langle L_z\rangle$, leading to an infinite initial slope (dotted
line in Fig.\ \ref{f4}). At fixed $\langle L_z\rangle$ and $k'_\mu$,
entanglement is then stronger in the unstable sector $B_1$
($\omega>\omega'_{c2}$). An {\it exceptional} behavior occurs in the critical
case $k'_y=-k'_x$ (Eq.\ (\ref{fcrit})), where for $\omega\rightarrow\infty$,
$\langle L_z\rangle\propto\omega^{-2}$ {\it vanishes} while $f_\mu$ remains
{\it finite}. In this special limit there is finite entanglement with {\it
vanishing} angular momentum.  On the other hand, close to the divergences,
$f_\mu\propto\langle L_z\rangle$ ($\omega\rightarrow \omega_c$) or $\langle
L_z\rangle^{1/2}$ ($\omega\rightarrow\omega'_{ci}$, $i=1,2$), implying
$S(\rho_\mu)\propto \ln\, \langle L_z\rangle$ for large $\langle  L_z\rangle$.

\subsection{Thermal entanglement}
\begin{figure}[t] \centering
\includegraphics[trim=0cm 0cm 0cm 0cm, width = 7cm]{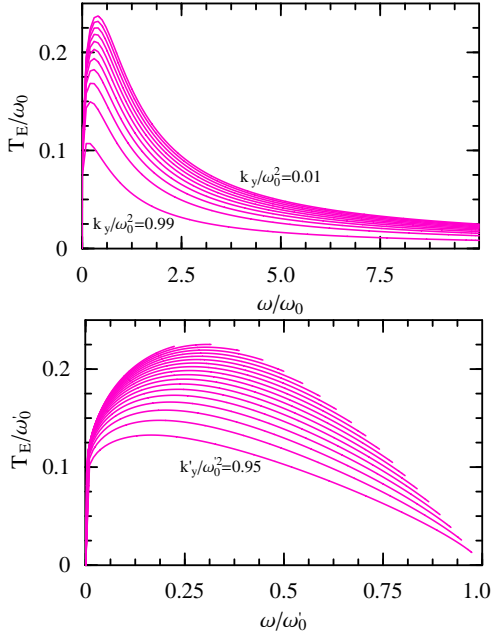}
\vspace*{-.75cm}

\caption{Top: Scaled limit temperatures for entanglement $T_E$
at fixed $k_\mu$ (top) and at fixed $k'_\mu$ (bottom), as a function of
frequency for different ratios $k_y/k_x=0.01,0.1,\ldots,0.9,0.99$ (top) and
$k'_y/k'_x=0.05,0.1,\ldots,0.95$ (bottom), with $\omega_0=\omega_x=\sqrt{k_x}$,
$\omega'_0=\omega'_x=\sqrt{k'_x}$. At fixed $k_\mu$, $T_E$ vanishes for large
$\omega$ (Eq.\ (\ref{TE2})) whereas at fixed $k'_\mu$, it approaches a finite
value at the upper stability limit $\omega'_{c1}=\sqrt{k'_y}$.} \label{f5}
\end{figure}

Let us now examine the thermal entanglement in the stable sector $A$. We first
depict in Fig.\ \ref{f5} the {\it limit temperature for entanglement} $T_E$,
determined from the condition $\tilde{f}_-=0$ (equality in Eq.\ (\ref{TE})).
This temperature {\it remains finite} for all values of $k_\mu$ or $k'_\mu$,
{\it including} the edge of the sector ($k_\mu\rightarrow 0$ or
$|\omega|\rightarrow\sqrt{k'_\mu}$), where the vacuum $f_\mu$ diverges. At the
edge, $\lambda_-\rightarrow 0$ and hence a finite $T$ already gives rise to a
spread over all energy levels ($f'_-\rightarrow\infty$), which diminishes and
eventually kills the entanglement. A related fundamental effect is that at {\it
finite} $T>0$, {\it  entanglement} {\it does
not diverge at the edge}, but stays finite or vanishes, depending on the value
of $T$.

More precisely, for $\omega_y\rightarrow 0$ and fixed $\omega_x>0$,
$\lambda_+\rightarrow \sqrt{4\omega^2+\omega_x^2}$ whereas  $\lambda_-\approx
\omega_y\omega_x/\lambda_+$, implying $f'_-\approx T/\lambda_-\approx
T\lambda_+/(\omega_x\omega_y)$. Hence, in this  limit Eqs.\
(\ref{fti})--(\ref{bef}) lead to
\begin{equation}
\tilde{f}_-=\frac{1}{2}[\sqrt{\frac{T\lambda_+^4(1+2f'_+)^2}
{\omega_x^2(2\omega^2\lambda_+(1+2f'_+)+T\omega_x^2)}}-1]
 \label{fta}\,,\end{equation}
which remains {\it finite and  above} $-1/2$ if $T>0$. This implies a {\it
finite negativity in this limit if $T>0$,} with $N\propto T^{-1/2}$ for
$T\rightarrow 0$ according to Eq.\ (\ref{Nt}). Therefore, at finite temperature
the vacuum divergences of the entanglement can be only probed indirectly,
through the $T^{-1/2}$ behavior of $N$ near the edge at sufficiently low $T$.

In addition, Eq.\ (\ref{fta}) entails a {\it finite limit temperature} $T_E$,
obtained from the condition $\tilde{f}_-=0$ in (\ref{fta}):
\begin{equation}
T_E={\frac{2(1+2f'_+)\omega^2\omega_x^2\lambda_+}{(1+2f'_+)^2\lambda_+^4-\omega_x^4}}\,,
 \label{TE1}\end{equation}
which is a transcendental equation for $T_E$ ($f'_+$ depends on $T_E$). The
{\it maximum limit temperature} $T_E^M$ at fixed $k_x=\omega_x^2$ or ${T'}_E^M$
at fixed $k'_x={\omega'_x}^2$, is in fact obtained in this limit ($\omega_y=0$
or $\omega'_y=\omega$): At fixed $k_x$, $T_E^M\approx 0.24\omega_x$, attained
at $\omega\approx 0.38\omega_x$, while at fixed $k'_x$, ${T'}_E^M\approx
0.23{\omega'_x}$, attained at $\omega\approx 0.28{\omega'}_x$.

 \begin{figure}[t] \centering
\includegraphics[trim=0cm 0cm 0cm 0cm, width = 7cm]{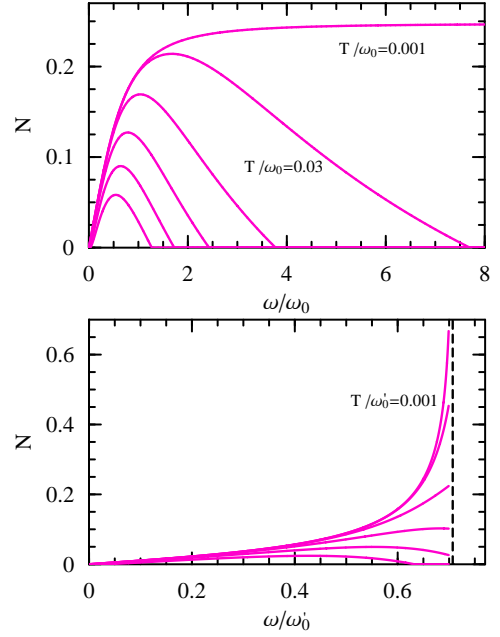}
\vspace*{-.75cm}

\caption{Top: Negativity as a function of frequency at increasing
temperatures $T/\omega_0=0.001,0.03,0.06,\ldots,0.15$, at fixed $k_\mu$ (top),
for $k_y/k_x=0.2$,  and at fixed $k'_\mu$ (bottom), for $k'_y/k'_x=0.5$. At $T>0$,
$N$ is finite, and non-zero just within a finite frequency window.} \label{f6}
\end{figure}

At fixed $k_\mu$, the limit temperature $T_E$ as a function of $\omega$
exhibits first a maximum and then {\it vanishes for  $\omega\rightarrow\infty$}
(top panel in Fig.\ \ref{f5}), i.e., in the limit where the vacuum entanglement
saturates. The reason is that $\lambda_-$ also vanishes for
$\omega\rightarrow\infty$ ($\lambda_-\approx \omega_x\omega_y/(2\omega)$),
implying  $T_E\propto \omega^{-1}$ in this limit:
\begin{equation}
T_E\approx \frac{\omega_x\omega_y}
{2\omega\ln\frac{\omega_x+\omega_y}{\omega_x-\omega_y}}\label{TE2}
\,.\end{equation}
In fact, for $\omega\rightarrow\infty$ and fixed $k_\mu$ (with $k_y<k_x$), Eqs.\
(\ref{fti})--(\ref{bef}) lead to
 \begin{equation}\tilde{f}_-\approx
 \frac{1}{2}[\sqrt{\frac{(1+2f'_-)\omega_y}{\omega_x}}-1]\,,\end{equation}
such that $\tilde{f}_-=0$ for $1+2f'_-=\omega_x/\omega_y$, which leads to Eq.\
(\ref{TE2}). On the other hand, for $\omega\rightarrow 0$, $T$ vanishes
logarithmically ($T_E\propto -1/\ln\omega$) and the same occurs for
$\omega_y\rightarrow\omega_x$, since in these limits $\lambda_\pm$ remain both
finite whereas the negativity vanish. At fixed $T$, we then obtain a {\it
finite frequency window for entanglement}, which narrows for increasing
temperature or decreasing anisotropy, as seen in Fig.\ 5 and also Fig.\
\ref{f6}, where the negativity (\ref{Nt}) is depicted. Let us remark that
entanglement ceases to be correlated with $\langle L_z\rangle$ as the
temperature increases ($\langle L_z\rangle\propto \omega T$ for high $T$).

At fixed $k'_\mu$, the behavior of $T_E$ and $N$ look quite different, as now
$\omega$ is bounded above by $\omega'_y$ (assuming $\omega'_y<\omega'_x$). For
$\omega\rightarrow\omega'_y$, $T_E$ is then determined by Eq.\ (\ref{TE1}) with
$\omega_x\rightarrow\sqrt{{\omega'_x}^2-\omega^2}$, and remains finite.
Actually, as verified in the bottom panel of Fig.\ \ref{f5}, $T_E$ acquires in
this border its maximum value as $\omega$ increases at fixed $k'_\mu$ if
$\omega'_y<\omega'_{yc}\approx 0.28\omega'_x$, while if
$\omega'_y>\omega'_{yc}$ the maximum is attained at an intermediate frequency.
Consequently, at fixed $T<{T'}_E^M$ there is again entanglement within a
certain frequency window, which extends up to the stability border
$\omega=\omega'_y$ if $\omega'_y<\omega'_{yc}$ or $T<T_E$ at $\omega=\omega'_y$
(bottom panel in Fig.\ \ref{f6}). The absolute maximum ${T'}_E^M$ is obtained
at this border precisely at $\omega'_y=\omega'_{yc}$.   For $\omega\rightarrow
0$ or $\omega'_y\rightarrow\omega'_x$, $T_E$ decreases again logarithmically.

\vspace*{-0.cm}
 \begin{figure}[t] \centering
\includegraphics[trim=0cm 0cm 0cm 0cm, width = 6cm]{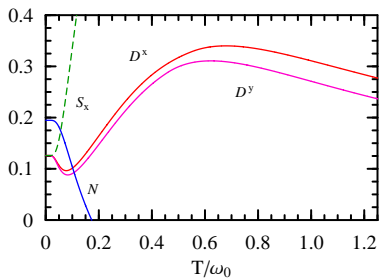}
\vspace*{-.25cm}

\caption{Negativity ($N$), single mode entropy $S_x=S(\rho_x)$
and the quantum discords $D^y$ and $D^x$ as a function of temperature for
$\omega=\omega_0=\sqrt{k_x}$ and $k_y=0.2 k_x$.}

\label{f7}
\end{figure}
\subsection{Comparison with the quantum discord}
We finally compare in Fig.\ \ref{f7} the thermal behavior of the negativity
with that of the gaussian quantum discord $D^y$ (Eq.\ (\ref{Dy})) and also
$D^x$. For reference we have also plotted the entropy of one of the modes
($x$), no longer a measure of entanglement, just to indicate its coincidence
with both $D^y$ and $D^x$ for $T\rightarrow 0$. While at $T=0$ the negativity
is just an increasing function of the entanglement entropy (Eqs.\
(\ref{ftvac})--(\ref{Smu})) and hence of the quantum discord, the behavior for
$T>0$ is quite different. Although exhibiting a similar initial decreasing
trend (essentially due to the initial increase of the total entropy
$S(\rho_{AB})$ in (\ref{DB})) the gaussian discord starts then to increase (due
to the increase in the first term of (\ref{DB})), vanishing only asymptotically
for $T\rightarrow\infty$. Such revival of the discord with increasing $T$ was
also observed in spin systems \cite{WR.10,MG.10}, and reflects the presence of
quantum correlations in the excited eigenstates, which lead at these
temperatures to a separable yet not classically correlated (in the sense of
sec.\ \ref{QDs}) thermal state. Since $D^\mu>1$ implies entanglement
\cite{GP.10,AD.10}, one can ensure here that $D^\mu<1$ after the vanishing of
the negativity ($T>T_E$), although this may not prevent $D^\mu$ from reaching a
higher value than at $T=0$ at some intermediate temperature, as seen in Fig.\
\ref{f7}. For $T\rightarrow\infty$  we actually obtain, from Eq.\ (\ref{Dy})
and the expression of \cite{AD.10}, that $D^\mu\propto T^{-1}$:
\begin{equation} D^y\approx \frac{\omega^2}{2T\sqrt{\omega^2+\omega_{x}^2}}
\,,\label{Dmu}
\end{equation}
with a similar expression for $D^x$ after replacing $\omega_x$ by $\omega_y$.
Hence, for high $T$ $D^\mu$ becomes independent of $\omega_{\mu}$, with
$D^x>D^y$ asymptotically if $\omega_y<\omega_x$, as verified in Fig.\ \ref{f7}.
We also remark that the discord remains finite for $T>0$ in the whole sector
$A$.

\section{Conclusions \label{IV}}
We have analyzed the entanglement induced by an angular momentum coupling on
two harmonic modes. Full analytic expressions for the vacuum entanglement
entropy and the thermal negativity were derived. The model exhibits a rich
phase structure and admits distinct physical realizations (particle in a
magnetic field in an anisotropic harmonic trap, or particle in a rotating
harmonic trap), which lead to different entanglement behaviors with the
relevant control parameter. For instance, in sector A (stable vacuum),
entanglement saturates for strong fields in the first case, but diverges at a
finite frequency in the second case. Vacuum entanglement diverges at the onset
of instabilities, being correlated with the average angular momentum and
reaching  higher values in unstable domains dynamically stabilized by the field
or rotation. In contrast, thermal entanglement is finite, and non-zero just
below a {\it finite} limit temperature within a reduced frequency window,
diverging only for $T\rightarrow 0$ at the instability borders.  We have also
shown that after a short initial common trend, the thermal behavior of the
gaussian quantum discord becomes substantially different from that of
entanglement, vanishing only asymptotically. A deeper investigation of the
discord and other related measures of quantum correlations \cite{AD.11,RCC.10}
in similar systems is being undertaken.

The authors acknowledge support from CONICET (LR) and CIC (RR) of Argentina.


\begin{thebibliography}{99}
\bibitem{NC.00}M.A.\ Nielsen and I.L.\ Chuang, {\it Quantum Computation and
        Quantum Information} (Cambridge Univ.\ Press, Cambridge, UK, 2000).
\bibitem{VE.07}V.\ Vedral, {\it Introduction to Quantum Information Science}
(Oxford Univ.\ Press, Oxford, UK 2006).
\bibitem{AFOV.08} L.\ Amico, R.\ Fazio, A.\ Osterloh and
           V.\ Vedral, Rev.\ Mod.\ Phys.\ {\bf  80}, 516  (2008).
\bibitem{HHHH.10} R.\ Horodecki, P.\ Horodecki, M.\ Horodecki and K.\ Horodecki, Rev.\ Mod.\ Phys.\ {\bf 81}, 865 (2009).
 \bibitem{RS.00} R.\ Simon, Phys.\ Rev.\ Lett.\ {\bf 84}, 2726 (2000).
\bibitem{DGC.00}L.M.\ Duan, G.\ Giedke, J.I.\ Cirac, P.\ Zoller,
Phys.\ Rev.\ Lett.\ {\bf 84}, 2722 (2000).
\bibitem{WW.01} R.F.\ Werner, M.M.\ Wolf, Phys.\ Rev.\ Lett.\ {\bf 86}, 3658 (2001).
 \bibitem{AEPW.02} K.\ Audenaert, J.\ Eisert, M.B.\ Plenio, and R.F.\ Werner,
 Phys.\ Rev.\ A {\bf 66}, 042327 (2002).
 \bibitem{ASI.04}G.~Adesso, A.~Serafini, and F.~Illuminati,
Phys.~Rev.~A {\bf 70}, 022318 (2004); A.~Serafini, G.~Adesso, and
F.~Illuminati, Phys.~Rev.~A {\bf 71}, 032349 (2005).
\bibitem{CEPD.06} M.~Cramer,  J.~Eisert, M.B.~Plenio, and J.~Drei\ss ig,
 Phys.\ Rev.\ A {\bf 73}, 012309 (2006).
\bibitem{AI.08} G.~Adesso, F.~Illuminati, Phys.~Rev.~A {\bf 78}, 042310 (2008).
\bibitem{MRC.10} J.M.\ Matera, R.\ Rossignoli, N.\ Canosa,
Phys.~Rev.~A {\bf 82}, 052332 (2010).
\bibitem{BBPS.96} C.H.\ Bennett, H.J.\ Bernstein, S.\ Popescu, and B.\ Schumacher, Phys. Rev. A {\bf 53}, 2046 (1996).
\bibitem{ZHSL.99}K.\ Zyczkowski, P.\ Horodecki, A.\ Sanpera, and M.\ Lewenstein, Phys. Rev. A {\bf 58,} 883 (1998).
\bibitem{VW.02} G.\ Vidal, R.F.\ Werner, Phys. Rev. A {\bf 65}, 032314 (2002).
\bibitem{PP.96}A.\ Peres, Phys.\ Rev.\ Lett.\ {\bf 77}, 1413 (1996).
\bibitem{HHH.96}M.\ Horodecki, P.\ Horodecki and R.\ Horodecki,
Phys.\ Lett.\ {\bf A 223}, 1 (1996).
\bibitem{BvL.05}S.L.\ Braunstein and P.\ van Loock, 
 Rev.\ Mod.\ Phys.\ {\bf 77}, 513 (2005).
\bibitem{OZ.01} H.\ Ollivier and W.H.\ Zurek, Phys.\ Rev.\ Lett.\ {\bf 88},
 017901 (2001).
\bibitem{HV.01} L.\ Henderson and V.\ Vedral, J.\ Phys.\ A {\bf 34}, 6899 (2001).
\bibitem{GP.10} P.\ Giorda, M.G.A.\ Paris, Phys.\ Rev.\ Lett.\ {\bf 105},
020503 (2010).
\bibitem{AD.10} G.\ Adesso, A.\ Datta, Phys.\ Rev.\ Lett.\ {\bf 105}, 030501 (2010).
\bibitem{DSC.08} A.\ Datta, A.\ Shaji, and C.M.\ Caves, Phys.\ Rev.\ Lett.\
{\bf 100}, 050502 (2008).
\bibitem{KL.98} E.\ Knill, R.\ Laflamme, Phys.\ Rev.\ Lett.\ {\bf 81},
5672 (1998).
\bibitem{DFC.05} A.\ Datta, S.T.\ Flammia and C.M.\ Caves, Phys.\ Rev.\ A {\bf 72}, 042316 (2005).
\bibitem{CAB.11} D.\ Cavalcanti et al, Phys.\ Rev.\ A {\bf 83}, 032324 (2011).
\bibitem{SKD.11} A.\ Streltsov, H.\ Kampermann, and D.\ Bru\ss,
Phys.\ Rev.\ Lett.\ {\bf 106}, 160401 (2011).
\bibitem{PGA.11} M.\ Piani et al, Phys.\ Rev.\ Lett.\ {\bf 106}, 220403 (2011).
\bibitem{Va.56} J.G. Valatin, Proc.\ R.\ Soc.\ London {\bf 238}, 132 (1956).
\bibitem{FK.70} A.\ Feldman and A.\ H.\ Kahn, Phys.\ Rev.\ B {\bf 1}, 4584 (1970).
\bibitem{RS.80} P.~Ring and P.~Schuck, {\it The Nuclear Many-Body Problem},
(Springer, NY, 1980).
\bibitem{BR.86} J.P. Blaizot and G. Ripka, {\it Quantum Theory of Finite
Systems} (MIT Press, MA, 1986).
\bibitem{MC.94}A.V.\ Madhav, T.\ Chakraborty, Phys.\ Rev.\ B {\bf 49}, 8163 (1994).
\bibitem{LNF.01} M.\ Linn, M.\ Niemeyer, and A.\ L.\ Fetter,
Phys.\ Rev.\ A {\bf 64}, 023602 (2001).
\bibitem{OO.04} M.\ \"O.\ Oktel, Phys.\ Rev.\ A {\bf 69}, 023618 (2004).
\bibitem{AF.07} A.\ L.\ Fetter, Phys.\ Rev.\ A {\bf 75}, 013620 (2007).
\bibitem{ABL.09} A.\ Aftalion, X.\ Blanc, and N.\ Lerner,
Phys.\ Rev.\ A {\bf 79}, 011603(R) (2009).
\bibitem{ABD.05} A.\ Aftalion, X.\ Blanc, J.\ Dalibard,
Phys.\ Rev.\ A {\bf 71}, 023611 (2005);
S. Stock et al, Laser Phys.\ Lett. {\bf 2}, 275 (2005).
\bibitem{BDS.08}
I.\ Bloch, J.\ Dalibard, W.\ Zwerger, Rev.\ Mod.\ Phys.\ {\bf 80}, 885 (2008).
\bibitem{AF.09} A.\L.\ Fetter, Rev.\ Mod.\ Phys.\ {\bf 81}, 647 (2009).
\bibitem{PE.94}J.\ P\v{e}rina, Z.\ Hradil, and B.\ Jur\v{c}o,
{\it Quantum optics and Fundamentals of Physics} (Kluwer, Dordrecht, 1994);
N.\ Korolkova, J.\ P\v{e}rina, Opt.\ Comm.\ {\bf 136}, 135 (1996).
\bibitem{RK.09} R.\ Rossignoli and A.M.\ Kowalski, Phys.\ Rev.\ A {\bf 79}
062103 (2009). 
\bibitem{HMM.03} A.P.\ Hines, R.\ H.\ McKenzie, and G.J.\ Milburn,
Phys.\ Rev.\ A {\bf 67}, 013609 (2003).
\bibitem{NL.05}H.T.\ Ng, P.T.\ Leung, Phys.\ Rev.\ A {\bf 71}, 013601 (2005).
\bibitem{CN.08} A.V.\ Chizhov, R.G.\ Nazmitdinov, Phys.\ Rev.\ A {\bf 78},
064302 (2008).
\bibitem{RW.89} R.F.\ Werner, Phys.\ Rev.\ A {\bf 40}, 4277 (1989).
\bibitem{FA.10} A.\ Ferraro et al, Phys.\ Rev.\ A {\bf 81}, 052318 (2010).
\bibitem{AD.11}L.\ Mi\v{s}ta et al, Phys.\ Rev.\ A {\bf 83}, 042325 (2011).
\bibitem{WR.10} T.\ Werlang and G.\ Rigolin, Phys.\ Rev.\ A {\bf 81},
044101 (2010).
\bibitem{MG.10} J.\ Maziero et al, Phys.\ Rev.\ A {\bf 82}, 012106 (2010).
\bibitem{RCC.10} R.\ Rossignoli, N.\ Canosa and L.\ Ciliberti,
Phys.\ Rev.\ A {\bf 82}, 052342 (2010).
\end{thebibliography}
\end{document}